\documentclass[aps,prb,superscriptaddress,twocolumn]{revtex4-2}

\usepackage{bm}
\usepackage{dsfont}
\usepackage{graphicx}
\usepackage{amsmath}
\usepackage{txfonts}
\usepackage{amssymb}
\usepackage{color}
\usepackage{subfigure}
\usepackage[UKenglish]{babel}

\renewcommand{\H}{{\cal{H}}}
\newcommand{\bS}{\bm{S}}
\newcommand{\bs}{\bm{s}}
\newcommand{\hti}{{\tilde{h}}}

\newcommand{\bh}{{\bf h}}
\newcommand{\tG}{t_{_{\rm G}}}

\newcommand{\muB}{\mu_{_{\rm B}}}

\begin{document}

\title{Full-magnetic implementation of a classical Toffoli gate}

\author{Davide Nuzzi}
\affiliation{Dipartimento di Fisica e Astronomia, Universit\`a di Firenze, Via G. Sansone 1, I-50019 Sesto Fiorentino (FI), Italy}
\affiliation{Istituto Nazionale di Fisica Nucleare, Sezione di Firenze, via G.~Sansone 1, I-50019 Sesto Fiorentino (FI), Italy}
\author{Leonardo Banchi}
\email{leonardo.banchi@unifi.it}
\affiliation{Dipartimento di Fisica e Astronomia, Universit\`a di Firenze, Via G. Sansone 1, I-50019 Sesto Fiorentino (FI), Italy}
\affiliation{Istituto Nazionale di Fisica Nucleare, Sezione di Firenze, via G.~Sansone 1, I-50019 Sesto Fiorentino (FI), Italy}

\author{Ruggero Vaia}
\email{ruggero.vaia@cnr.it}
\affiliation{Istituto dei Sistemi Complessi, Consiglio Nazionale delle Ricerche, via Madonna del Piano 10, I-50019 Sesto Fiorentino (FI), Italy}
\affiliation{Istituto Nazionale di Fisica Nucleare, Sezione di Firenze, via G.~Sansone 1, I-50019 Sesto Fiorentino (FI), Italy}

\author{Enrico Compagno }
\affiliation{ Department of Physics and Astronomy, University College London, Gower St., London WC1E 6BT, UK}
\affiliation{CNRS, Institut NEEL, F-38042 Grenoble, France}
\affiliation{Universit\'{e} Grenoble-Alpes, Institut NEEL, F-38042 Grenoble, France}

\author{Alessandro Cuccoli}
\affiliation{Dipartimento di Fisica e Astronomia, Universit\`a di Firenze, Via G. Sansone 1, I-50019 Sesto Fiorentino (FI), Italy}
\affiliation{Istituto Nazionale di Fisica Nucleare, Sezione di Firenze, via G.~Sansone 1, I-50019 Sesto Fiorentino (FI), Italy}

\author{Paola Verrucchi}
\affiliation{Istituto dei Sistemi Complessi, Consiglio Nazionale delle Ricerche, via Madonna del Piano 10, I-50019 Sesto Fiorentino (FI), Italy}
\affiliation{Dipartimento di Fisica e Astronomia, Universit\`a di Firenze, Via G. Sansone 1, I-50019 Sesto Fiorentino (FI), Italy}
\affiliation{Istituto Nazionale di Fisica Nucleare, Sezione di Firenze, via G.~Sansone 1, I-50019 Sesto Fiorentino (FI), Italy}

\author{Sougato Bose}
\affiliation{ Department of Physics and Astronomy, University College London, Gower St., London WC1E 6BT, UK}

\date{\today}

\begin{abstract}
	The Toffoli gate is the essential ingredient for reversible computing, an energy efficient 
	classical computational paradigm that evades the energy dissipation resulting from Landauer's principle. 
	In this paper we analyze different setups to realize a magnetic
	implementation of the Toffoli gate using three interacting classical spins,
	each one embodying one of the three bits needed for the Toffoli gate. 
	In our scheme, different control-spins configurations produce 
	an effective field capable of conditionally flipping the target spin. 
	We study what are the experimental requirements for the realization of our scheme, focusing on the 
	degree of local control, the ability to dynamically switch the spin-spin
	interactions, and the required single-spin anisotropies to make the classical
	spin stable, showing that these are compatible with current technology. 
\end{abstract}

\maketitle

\section{Introduction}
\label{s.basics}
Reversible computing is a computational paradigm inspired by physical aspects, where 
the elementary logic operations  and logic circuits are ideally invertible \cite{Toffoli1982}.
The main motivation for reversible computation is to minimize the energy costs   
by avoiding the requirement of logic operations whose number of outputs is lower than the number of 
inputs. Indeed, due to the Landauer's principle \cite{Bennett2003}, these irreversible operations 
dissipate energy, and therefore limit the efficiency of the resulting 
computation. A central result in reversible computing is that {\it any} algorithm can be 
realised through the use of a single operation called Toffoli gate \cite{Toffoli1982,Fredkin1982}, a three-bit to three-bit Control-Control-NOT operation, where the value of the target bit is reversed depending on the value of the first two bits. It is also known that the quantum version of the Toffoli gate is  universal for quantum computing, when paired
with the Hadamard gate
\cite{Shi2003} -- quantum computing is ideally reversible by physical principles, 
given the unitarity of quantum evolution. It has been shown that 
a quantum Toffoli gate can be achieved using the evolution of quantum spin systems 
either with external control pulses \cite{Sanders2015} or via unmodulated interactions 
\cite{Banchi2016}. 
On the other hand, because of the reversibility of quantum evolution, it has been proved that even a classical Toffoli 
gate can be obtained from the physical dynamics of quantum spins \cite{AntonioRHMB2015}. 
An implementation of the classical Toffoli gate using a continuous-time machine was proposed 
by Toffoli \cite{toffoli1981bicontinuous}, which was later extended to the quantum setting by 
Feynman \cite{feynman1985quantum}. 

In this paper we aim at the possibility of realising a classical Toffoli gate exploiting the continuous-time dynamics of interacting classical spins. A classical spin corresponds to the limit of a quantum spin for large spin quantum number $S$ and is represented as a fixed-modulus vector in a three-dimensional space. The building blocks of classical computing with interacting large-$S$ spins have recently been explored in experiments with magnetic adatoms \cite{KhajetooriansWCW2011}, using irreversible operations and thermal equilibrium states that result from the interaction with the substrate environment. On the other hand, in this work we are interested in the reversible dynamics of classical spins in a low relaxation limit, a regime that can be obtained, for instance, in molecular systems \cite{GatteschiSessoliVillain06}.
More recently, Toffoli gates were implemented using magnetic skyrmions \cite{Toffoli2023}, 
silicon spin qubits \cite{Gullans2019}, and a setup involving 
spin waveguides, cross-junctions, and phase shifters \cite{Balynskiy2018}. 

The dynamics of classical spins is ruled by Hamiltonians and Poisson brackets that follow from the spin commutation rules in the large-$S$ limit, so they can be imagined as objects carrying a fixed angular momentum and realized as magnetic dipoles.


   
In the following, we present two different models, whose common feature is the presence of three interacting classical spins, each one embodying one of the three bits needed for the Toffoli gate. In other words, the two logical states of the bits are encoded in the alignment (1-state) or anti-alignment (0-state) of the corresponding spin along a locally preferred direction. The basic idea behind our schemes is that different configurations of the two control-spins give rise to different effective magnetic fields acting on the target spin, thus producing different dynamics. 

The two models differ for the choice of the local easy-axis direction: in the first one, introduced in Section~\ref{s.coll.model}, all spins, being they either control or target, are supposed to be collinear when in their stable steady states, while in the second model they are taken to form mutual well defined angles, as detailed in Section~\ref{s.noncoll.model}. In each Section we discuss the role played by single-ion anisotropy and dissipation. Finally, to better frame the problem in view of a concrete realization, a quantitative estimate of the relevant parameters is given in the final Section.



\section{Collinear-spin model }
\label{s.coll.model}

A quantum spin corresponds to an angular momentum operator $\hat{\bS}$ having a fixed modulus, $\hat{\bS}^2=(\hbar{\cal{S}})^2$, ${\cal{S}}$ being the (semi-integer) spin quantum number. Its three components obey the commutation relations
\begin{equation}
 \big[\hat{S}^\alpha,\hat{S}^\beta\big]
 =i\hbar\varepsilon^{\alpha\beta\gamma}\hat{S}^\gamma ~,
\label{e.spinCR}
\end{equation}
where Greek letters correspond to the components $x$, $y$, and $z$, and $\varepsilon^{\alpha\beta\gamma}$ is the completely antisymmetric Levi-Civita symbol.
The classical limit $\hbar\to{0}$ entails that the spin quantum number ${\cal{S}}\to\infty$ in such a way that the modulus $\hbar{\cal{S}}\to{S}$ is kept constant: $\hat{\bS}$ becomes a classical vector $\bS$, i.e., with three commuting components. Being its modulus $|\bS|=S$ a constant (with the dimensions of an action), it is more convenient to deal with the unit vector $\bs=\bS/S$, whose components obey the Poisson brackets that follow from Eq.~\eqref{e.spinCR},
\begin{equation}
 \Big\{s^\alpha,s^\beta\Big\}=S^{-1}\varepsilon^{\alpha\beta\gamma}\,s^\gamma ~.
\label{e.spinPB}
\end{equation}

Our aim is to define a spin system and its dynamics in such a way that it can implement classical logical operations. To accomplish this goal we first specify how a classical spin can be exploited to encode a classical {\it bit}. If the spin is subjected to an easy-axis anisotropy that favors its alignment along, say, the $z$-direction, it can have two stable directions, $s^z=\pm{1}$. The natural mapping is to encode the classical bit into two stable orientations,
\begin{equation}
 \text{bit states } \{0, 1\} 
 \quad \Longleftrightarrow \quad
 \bs = \{\,(0,0,-1),\, (0,0,1)\,\}~,
\label{e.logicstates}
\end{equation} 
that may be dubbed South- and North pole.

\smallskip
As our purpose is to single out a dynamics that could implement a three-bits Control-Control-Operation, here we consider a three classical spins system: $\bs_1$ and $\bs_2$ are {\em control} spins while $\bs\equiv(x,y,z)$ is the {\em target} spin. Besides the easy-axis anisotropy, we  consider an exchange interaction between control and target spin, while adjustable magnetic fields are introduced as tunable parameters for driving the dynamics in a controlled way. The Hamiltonian of the system is 
\begin{equation}
	\H = -JS^2\Big[( s_1^z {+} s_2^z) z 
         +a \big( {s_1^z}^2 {+} {s_2^z}^2 {+} z^2 \big) 
         + h_\|\big( {s_1^z} {+} {s_2^z} {+} z \big)
         + h_\perp x\Big] ~,
\label{e.ham}
\end{equation}
where the exchange interaction between the control spins and the target spin is  ferromagnetic, $J>0$, and of the Ising type. Hereafter $JS^2$ is  used as the energy unit and $(JS)^{-1}$ as the time unit. The single-site easy-axis anisotropy $a>0$ defines the $z$-direction and the stable up/down configurations. Finally, an overall Zeeman field $h_\|$ along the $z$-direction is applied, while a local tunable field $h_\perp$ (one can assume it to be non-negative, without loss of generality) along the $x$ direction is assumed to act locally on the target spin.

The equations of motion (EoM) are obtained from the Poisson brackets~\eqref{e.spinPB} through $d(\cdot)/{dt}\,=\,\big\{(\cdot),\H\big\}$ and the specific form of the Hamiltonian~\eqref{e.ham} implies that $\dot{s}_1^z=\dot{s}_2^z=0$, namely, the $z$-component of the control spins is conserved. This implies that, if initially in $s_i^z=\pm1$, the control spins do not evolve in time. This is a desirable feature to implement a Control-Control-Operation, such as the Toffoli gate, which is defined by the following rule: if and only if both control spins are in the $1$ configuration (with the encoding~\eqref{e.logicstates}) the target spin is flipped (from 0 to 1 or vice-versa), otherwise it is unchanged.

For the dynamics of the target spin, the Hamiltonian~\eqref{e.ham} yields the following EoM, 
\begin{equation}
 \dot\bs = \bs \times \bh~,
\label{e.mag_eom}
\end{equation} 
with effective field
\begin{equation}
 \bh = \big(h_\perp,~0,~h_\| + s_1^z + s_2^z + 2a z \big)~.
\label{e.eff_field}
\end{equation}
If $\bh$ is constant throughout the target spin motion (namely, if $a=0$), then the above equations describe the precession of $\bs$ around the direction identified by $\bh$, which depends on the  constant control-spin components $s_1^z$ and $s_2^z$. Thus, it is necessary to analyze the dynamical behavior of $\bs$ starting from the four different initial configurations of the control spins/bits, $[\bs_1~ \bs_2]$, according to the encoding of Eq.~\eqref{e.logicstates}. 
When $a\neq0$ Eq.~\eqref{e.mag_eom} is nonlinear, the effective field varies with $\bs$ and the dynamics is no more a simple precession. In the following section we first consider the case $a=0$.

\subsection{No single-ion anisotropy}

The case of vanishing anisotropy elucidates the mechanism for the realization of the Toffoli gate via the spin dynamics. For $a=0$ the EoM~\eqref{e.mag_eom} describe a precession of Larmor frequency $|\bh|$ around the fixed axis $\bh$. According to Eq.~\eqref{e.eff_field}, the different precession axes for the possible values of the controls are
\begin{equation}
\begin{aligned}
              \bh_{11} =&~ \big(~ {h}_\perp ,~0,~ 2 + {h}_\| ~\big)~,\\
 \bh_{10} = \bh_{01} =&~ \big(~ {h}_\perp ,~0,~ {h}_\| ~\big)~,\\
              \bh_{00} =&~ \big(~ {h}_\perp ,~0,~ -2 + {h}_\| ~\big) ~,
\end{aligned}
\label{e.configfield}
\end{equation}
so that fixing the values of ${h}_\|$ and ${h}_\perp$, different trajectories are obtained depending on the initial configuration of the control spins. 

A Toffoli gate requires that the dynamical evolution, at time $\tG$, defined as the {\it gate time}, produces the transformation on the target spin
\begin{equation}
 \bs(t{=}0) = (0,0,\pm1) \left\{~~
\begin{aligned}
  \xrightarrow[]{~~~~~[11]~~~~~~}~&~\bs(\tG)=(0,0,\mp1)
\\
  \xrightarrow[]{[00]~[01]~[10]}   ~&~\bs(\tG)=(0,0,\pm1)~. 
\end{aligned} \right.
\label{e.toff_output}
\end{equation}
The first action on $\bs$ can be achieved only if the precession axis $\bh_{11}$ lies in the $xy$-plane. From the first of Eqs.~\eqref{e.configfield}, this is obtained with the choice 
\begin{equation}
 {h}_\| = -2~, 
\label{e.xcond}
\end{equation}
while ${h}_\perp$ remains the only free parameter that sets the angular frequencies in the different cases. Here it is important to note  that the direction of the precession axis, when the control spins are different from the configuration [11], is not relevant since the second condition requires the target to come back to the initial configuration. The fundamental aspect is that the different conditions should be matched at the same gate time $\tG$. This request imposes constraints on the precession periods for the different control configurations, which are
\begin{equation}
\begin{aligned}
 &T_{00} = \frac{ 2 \pi }{\sqrt{h_\perp^2 + 16}}~, 
 &T_{11} = \frac{ 2 \pi }{h_\perp}~,
\\
 &T_{01} = T_{10} = \frac{ 2 \pi }{\sqrt{h_\perp^2 + 4}}~.
\end{aligned}
\label{e.periods}
\end{equation} 
In order to obtain the actions described in~\eqref{e.toff_output}, $\tG$ must be an odd multiple of $T_{11}/2$ and an integer multiple of $T_{01}$ and $T_{00}$, i.e.,
\begin{equation}
	\tG = \frac{2n+1}2T_{11} = m T_{01} = l T_{00}~,
\label{e.tg_cond}
\end{equation}
for three integer numbers $n,~m,~l$.
Comparing the third and the fourth term with the second one, it must be 
\begin{equation}
 \sqrt{1+4\,h_\perp^{-2}} = \frac{2m}{2n+1}
~,~~~
 \sqrt{1+16\,h_\perp^{-2}} = \frac{2l}{2n+1}~,
\label{e.condit_comp1}
\end{equation}
and the elimination of $h_\perp$ leads to a constraint for the three integers
\begin{equation}
 16\,m^2-4\,l^2 = 3(2n+1)^2 ~.
\label{e.condit_comp2}
\end{equation} 
The latter condition cannot be satisfied, as the left and right terms are even and odd, respectively. Hence, there is no choice of the parameters that allows to exactly achieve the desired transformation, Eq. \eqref{e.toff_output}.

However, this also suggests an interesting comparison: the condition~\eqref{e.condit_comp2} is exactly the same found in Ref.~\onlinecite{AntonioRHMB2015}, where, using a three-qubit Hamiltonian and quantum dynamics, the problem of realizing a classical Toffoli gate was afforded. In particular, comparing the Hamiltonian~\eqref{e.ham} (with $a=0$) with Eq.~(1) of Ref.~\onlinecite{AntonioRHMB2015} a precise correspondence is found between Eqs.~(5) in~\onlinecite{AntonioRHMB2015} and our Eqs.~\eqref{e.condit_comp1}, where the condition~\eqref{e.xcond} (equivalent to $\omega_2=2J_{zz}$ in Ref.~\onlinecite{AntonioRHMB2015}) is also enforced.
Therefore, the constraints on the Hamiltonian parameters required for obtaining a Toffoli gate are exactly the same in the pure quantum case (using qubits) and in the classical case (using classical spins). This is not so surprising, if one recalls the identity between the classical equations of motion (\ref{e.mag_eom}) and the quantum Heisenberg equations for spin operators interacting with an applied field. Such correspondence allows us to follow the reasoning of Ref.~\onlinecite{AntonioRHMB2015} to obtain an {\it approximate} Toffoli gate when $a=0$. In fact, for $n=0$ (for the lowest possible $\tG$) and for a large value of $m$ (i.e. $h_\perp=(m^2-1/4)^{-1/2}\sim 1/m$), assuming exactly valid the first of the~\eqref{e.condit_comp1}, the second equation gives $l=2m+{\cal{O}}(m^{-1})$, which is the closer to an integer the larger the value of $m$.

\begin{figure}[tb]
\includegraphics[width=.47\textwidth]{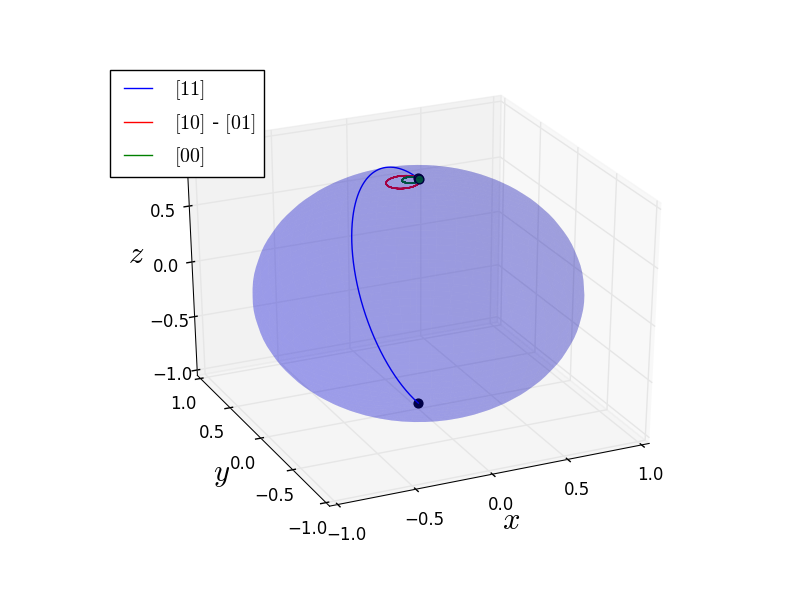}
\caption{Target spin $\bs$ trajectories on the unit sphere with $h_\|=-2$ and $h_\perp=0.201$ (i.e. the value obtained from the first of Eqs.~\eqref{e.condit_comp1} with $n=0$ and $m=5$). As shown in the legend panel, different colors correspond to different control configurations.}
\label{f.ex_appr_toff}
\end{figure}

Since one can weaken the encoding rule of the spin/bits, allowing them to represent $1$ or $0$ if their third component's modulus exceeds a certain threshold value instead of being exactly one (i.e., they do not have to exactly point the poles, but they have to stay in small finite regions close to them), it is possible to exploit the above reasoning to achieve the desired gate using a value of $m$ large enough to meet the second of the conditions~\eqref{e.condit_comp1} within the necessary precision. This is shown for example in  Fig.~\ref{f.ex_appr_toff}.
Moreover, the external environment can provide relaxation processes that enhance the stability of these regions, as discussed in Section~\ref{s.damping}, which is particularly important for applications when the device is used multiple times to realize many gates. 

\subsection{Single-ion anisotropy}
\label{s.aniso}

In this section we analyse the dynamics of the target spin $\bs$ for a finite value of the anisotropy parameter (i.e., for $a\neq0$). In this case, the motion described by Eqs.~\eqref{e.mag_eom} is not a simple precession, as the effective field $\bh$, which depends on the value of $s^z$, changes during the motion. 

As there are no simple analytical solutions in this case, we integrate numerically the EoM. Specifically, we analyze the dynamics of the target $\bs$ to find a single gate time $\tG$ which accomplishes the transformation ~\eqref{e.toff_output}, exploiting the freedom given by the three parameters $h_\perp$, $h_\|$, and $a$.

The general expressions are derived in Appendix~\ref{app}, and assume a simpler form for $\hti\,{=}\,0$, namely when the controls are in the state [11] and $h_\|$ is given by Eq.~\eqref{e.xcond}. In this case (see Eq.~\eqref{e.first_int} for the general case), the function $z(t)$ satisfies  
\begin{equation}
  {\dot z}^2 = (1-z^2)\,\big[h_\perp^2-a^2(1-z^2)\big]~,
\end{equation}
and setting $z=\cos\theta$, it further simplifies to
\begin{equation}
 {\dot\theta}^2=h_\perp^2-a^2 \sin^2\theta~.
\label{e.sz2_simpl_eq}
\end{equation}

As the condition $a|\sin\theta\,|\le{h_\perp}$ must be satisfied, it follows that for an anisotropy value larger than the field strength ($a>{h_\perp}$) the target spin cannot cross the equator ($\sin\theta=1$) and its motion is  confined to a cone of aperture $\sin^{-1}(h_\perp/a)$ around the pole where its dynamics started, $\theta(0)=0$ or $\theta(0)=\pi$. Therefore, the condition $a<{h_\perp}$ must be fulfilled for achieving a target spin to flip between the poles. Under this condition, the implicit kinematic equation reads 
\begin{equation}
 t=\int\limits_{0,\pi}^{\theta(t)} \frac{du}{\sqrt{h_\perp^2-a^2 \sin^2u}}~.
\label{e.sz2_simpl_keq}
\end{equation}
The resulting trajectories are closed and connect the two poles in a time corresponding to half-period of the motion,
\begin{equation}
 \frac{T_{[11]}}2 
  = \frac{\pi}{h_\perp}K\left(\frac{a}{h_\perp}\right)~,
\label{e.gatetime}
\end{equation}
where 
\begin{align}
	K(k) \equiv \frac1\pi\int_0^{\pi} \frac{du}{\sqrt{1-k^2\sin^2(u)}}
         = 1 + \frac{k^2}4 + \frac{9\,k^4}{64} + ... ~,
	\label{e.K}
\end{align}
is the complete elliptic integral of the first kind. Therefore,  the effect of anisotropy is to increase the precession period $T_{11}=2\pi/h_\perp$ in Eqs.~\eqref{e.periods}. Note that if the condition~\eqref{e.xcond} were not satisfied, the trajectory of $\bs$ would not cross the opposite pole, since in that configuration Eq.~\eqref{e.first_int} in both cases becomes ${\dot{z}}^2=-4\hti^2$.

\begin{figure}[t]
	\centering
	\includegraphics[width=\linewidth]{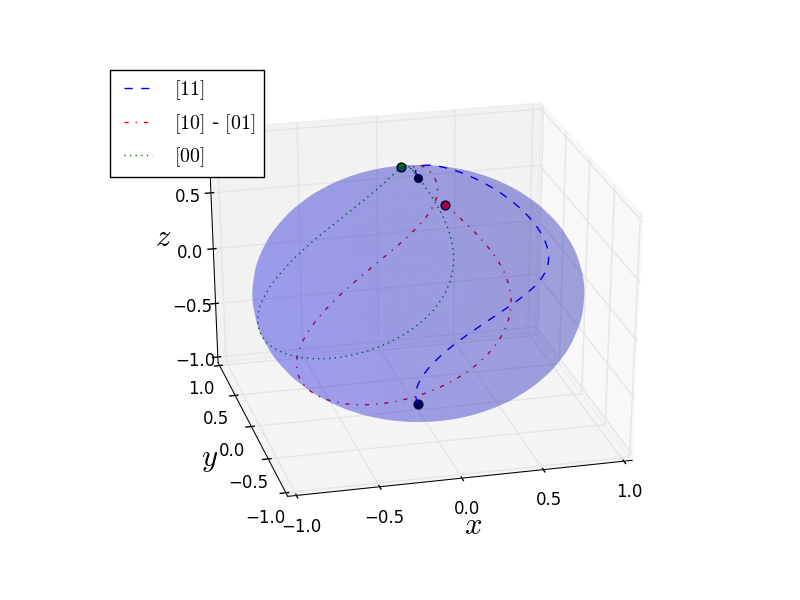}
  \caption{$\bs$ trajectories on the unit sphere with $h_\|=-2$, $h_\perp=2.7$ and $a=2.5$, 
		starting from the north pole. Final time is obtained from Eq.~\eqref{e.gatetime}. 
	Different colors correspond to different control configurations, as reported in the legend.}
	\label{f.anisofig}
\end{figure}

Because the motion is always periodic, there are time periods for any control-spin configuration (in particular those different from [11], where $\hti\ne0$) after which the target spin returns to the starting position. Then, as described before, a Toffoli gate occurs at time $\tG$ if Eqs.~\eqref{e.tg_cond} are satisfied. However, as analytic expressions for $T_{00}$ and $T_{10}$ are not available, we consider a numerical solution of Eq.~\eqref{e.impl_sol} in those cases. In Fig.~\ref{f.anisofig} the dynamics in presence of a non-zero anisotropy term is shown as an example.

 On the other hand, in Section~\ref{s.noncoll.model} we will show that the use of non-collinear control spins simplifies Eq.~\eqref{e.tg_cond} in the anisotropic case, since all the periods $T_{\alpha\beta}$ for any $\alpha,\beta=0,1$ can be expressed in terms of elliptic integrals. Although the exact solution of \eqref{e.tg_cond} may be impossible even in the presence of anisotropy,  we show in the following that approximate solutions are enough in the presence of Gilbert damping, since the latter stabilizes the evolution.

\subsection{Stability improvement due to external environments }
\label{s.damping}

Without an external environment, the dynamics described in the previous sections continues indefinitely so the driving field $h_\perp$ needs to be switched off externally to stop the dynamics and achieve the gate. As this scheme requires a careful timing of the switch control, its practical realization may be difficult. In the following we show that this limitation can be overcome by exploiting the interaction of the system with an external environment which provides a mechanism to stabilize the dynamics and help the realization of the Toffoli gate. The key point is that in a \textit{bistable} configuration  there are two stable minimum-energy spin orientations (North and South poles). The effect of the dissipation is then to drive the system to the ``closest'' equilibrium position, so that even if the Toffoli gate is implemented with some errors, as long as the imperfect dynamics brings the target spin close to the desired configuration, then relaxation processes transform such a dynamics into a perfect gate. In the following we show how this condition can be achieved. 

A damping term, $-\eta\,\bs\,{\times}\,\dot\bs$, that accounts for the energy loss of the magnetic moment caused by its interaction with the surroundings, is introduced following the scheme of the Laundau-Lifshitz-Gilbert (LLG) model. The equation~\eqref{e.mag_eom} for the target spin is then extended to 
\begin{equation}
 \big(1{+}\eta^2\big)\,\dot\bs = \bs \times\big[\bh - \eta\,(\bs\times\bh)\big]
                       \equiv {\bm L}(\bs)~.
\label{e.LLG_eq}
\end{equation}
We note that both the North and the South poles, $z=\pm1$, are equilibrium configurations for the LLG equation, when $h_\perp=0$. To check for the stability of these configurations we first linearize Eq.~\eqref{e.LLG_eq} around these points and we calculate the eigenvalues of the resulting matrix, $\big[\partial_\alpha L_\beta\big]_{z=\pm1}$, which are
\begin{equation}
 0
~,~~~~
 -(2a{+}z\hti)\,(\eta\pm i)~.
\end{equation}
Both poles are stable if the real part of the above eigenvalues is negative, i.e., $2a-|\hti|>0$. Therefore, taking into account the condition ~\eqref{e.xcond}, that entails $|\hti|\le{4}$, the stability is guaranteed for an anisotropy strength that satisfies $a>2$. Moreover, the field $h_\perp$ has to be larger than $a$ to allow a spin flip of the target when the controls are in the configuration [11]. Therefore, the condition that allows to obtain the Toffoli gate with stable periodic orbits is 
\begin{equation}
	h_\perp>a>2~.
\end{equation}

\begin{figure*}[t]
\centering
\includegraphics[width=0.9\linewidth]{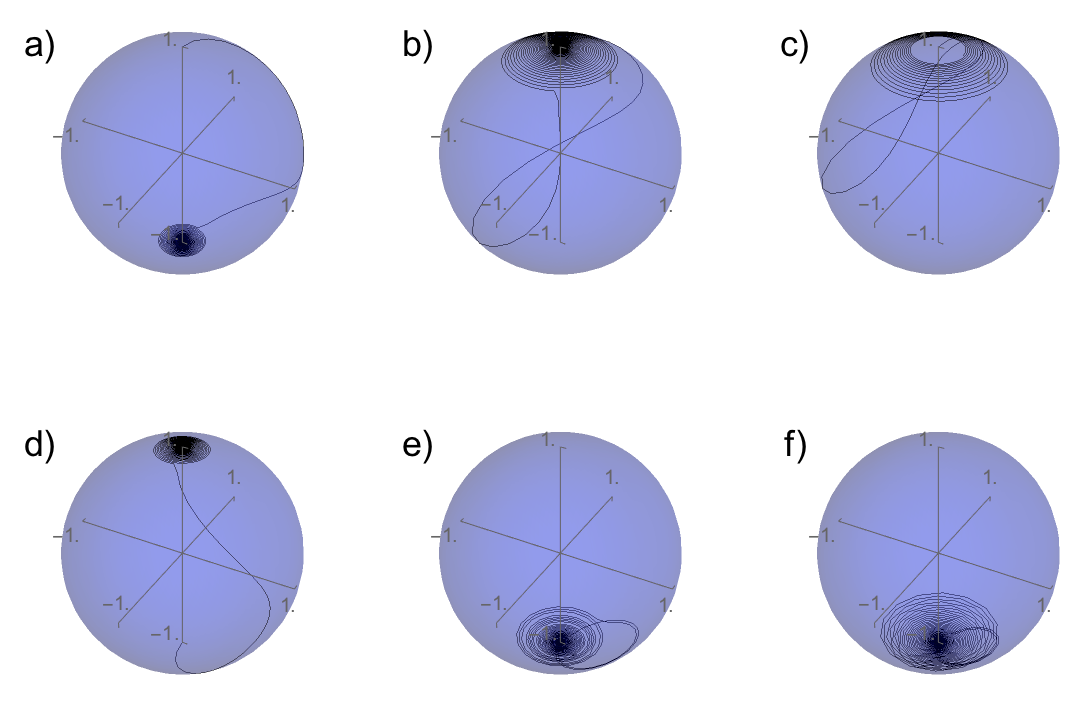}
\caption{Dissipative dynamics of the target spin following the LLG equation~\eqref{e.LLG_eq} for two different initial configurations along the $z$ axis, $s_T^z=1$ for the top row (a,b,c) and $s_T^z=-1$ for the bottom row (d,e,f), and different choices of the control spin, namely [11] for the first column (a,d), [00] for the central column (b,e) and [10] (equal to [01]) for the last column (c,f). 
	We used $a \simeq 2.5$, $h_\perp\simeq 2.7$ and $\eta=0.01$. 
The field $h_\perp$ is switched off after a time given by Eq.\eqref{e.gatetime}. 
}
\label{f.damped}
\end{figure*}
The effect of the damping factor for the dynamics of the target spin is plotted in Fig.~\ref{f.damped}, where it can be clearly seen that in all possible configurations of the control spins the target spin always converges to the desired pole. The damping term is assumed to be weak and the flipping time is found from Eq.~\eqref{e.gatetime}. This represents an estimate of the time in which the driving field, $h_\perp$, must be switched off to allow the relaxation of the target spin towards the closest equilibrium configuration. Even though the switching time is not perfectly matched, and even if the times $T_{\alpha\beta}$ do not exactly satisfy Eq.~\eqref{e.tg_cond}, one can see from Fig.~\ref{f.damped} that the dynamics implements a Toffoli gate that becomes exact in the long time limit, due to the sequential effects of the strong driving pulse and the stabilizing environment.

\section{Non-collinear-spin model}
\label{s.noncoll.model}

We now consider a different scenario where the stable positions of the control spins may be along different axes $\bm{e}_1$ and $\bm{e}_2$, while the target bit is defined in the $z$ direction. The values 1 and 0 of a control bit correspond to the spin being aligned or antialigned with the direction $\bm{e}_i$, i.e., $\bs_i=\bm{e}_i\Leftrightarrow{1}$ and  $\bs_i=-\bm{e}_i\Leftrightarrow{0}$. 
To simplify the analysis, the control spins are assumed to be frozen in their state, while the effect of the anisotropy is taken into account in the next subsection. 

Unlike the Ising interaction considered in Eq.~\eqref{e.ham}, here the target spin $\bs$ has a Heisenberg coupling with the control spins and is subjected to an external field $\bm{h}$,
\begin{equation}
 \H=(\bs_1+\bs_2+\bm{h})\cdot\bs~.
 \label{H.noncollinear}
\end{equation}
%
Since the control spins $\bs_1$ and $\bs_2$ are frozen, the resulting dynamics
is a precession around the effective field $\bs_1+\bs_2+\bm{h}$ with frequency
$|\bs_1+\bs_2+\bm{h}|$. It is easy to see that the choice of 
$\bm{h}=\bm{e}_1+\bm{e}_2$ freezes the dynamics for the [00] configuration, for which
$\H_{00}=0$ or, equivalently, the precession frequency is
$\omega_{00}=0$. On the other hand,
\begin{align}
	\H_{10}&=2\,\bm{e}_1{\cdot}\,\bs~,
     & \H_{01}=2\,\bm{e}_2\cdot\bs ~,
\end{align}
respectively for the configurations [10] and [01] so that $\omega_{10}=\omega_{01}=2$, namely the target spin comes back to the initial state at time $T_{[01]}=\pi$ and for every integer multiple of it. As in the previous scheme, we define the Toffoli gate time 
at one of these times, namely $\tG=mT_{[01]}$, and for the [11] configuration, where
\begin{equation}
 \H_{11}=2(\bm{e}_1+\bm{e}_2)\cdot\bs
\end{equation}
and $\omega_{11}=2\,|\bm{e}_1{+}\,\bm{e}_2|$, we require that at $\tG$ the target spin be flipped. The gate time has to satisfy Eq.~\eqref{e.tg_cond}, namely $\tG=mT_{01}=\frac{2n+1}2\,T_{11}$. The latter can be rewritten as $\omega_{11}=\omega_{10}\,\frac{2n+1}{2m}$ or, equivalently, 
\begin{equation}
 |\bm{e}_1+\bm{e}_2| \equiv 2\cos\varphi = \frac{2n+1}{2m}~,
\end{equation}
where $2\varphi$ is the angle between $\bm{e}_1$ and $\bm{e}_2$. Many solutions are possible (provided that the condition $2n+1<4m$ is satisfied). This amounts to require $\cos\varphi=\frac{2n+1}{4m}$, so that the simplest one is for $n=1$ and $m=1$, namely 
\begin{equation}
 \cos\varphi = \frac34
~~~ \Longrightarrow ~~~
 2\varphi \simeq 0.46\,\pi \simeq 83^\circ
\end{equation}

\subsection{Easy-axis anisotropy}
We consider the effect of a non-zero anisotropy $a$ for the target spin $\bs$ along the $z$-axis, $\H=(\bs_1+\bs_2+\bm{h})\cdot\bs + a(s^z)^2~.$ The equation of motion for the target spin $\bs$ reads
\begin{equation}
	\dot\bs = \bs \times (\bs_1+\bs_2+\bm{h} + 2az\bm{e}^z)~.
\end{equation}
We assume that now $\bm{e}_1$ and $\bm{e}_2$ lie in the $xy$-plane and we study the dynamics of the target spin, for all the control configurations, setting the $x$-axis in the direction of the in-plane field $\bs_1+\bs_2+\bm{h}\equiv(\hti,0,0)$, with $\hti_{00}=0$, $\hti_{10}=2$, $\hti_{01}=2$, and $\hti_{11}=2|\bm{e}_1+\bm{e}_2|=4\cos\varphi$. In this case the EoM can be cast in a form equivalent to Eq.~\eqref{e.sz2_simpl_eq}, where $h_\perp$ is replaced by $\hti$. For any control configuration $[\alpha\beta]$ one can use Eq.~\eqref{e.gatetime} to get the half period, for $\hti>a$, as 
\begin{equation}
	T_{[\alpha\beta]} = \frac{2\pi}{\tilde h_{\alpha\beta}}
		K\left(\frac{a}{\hti_{\alpha\beta}}\right)~.
	\label{e.gatetime.noncol}
\end{equation}
Similarly to what happened in the collinear case, from Eq.~\eqref{e.K} we find that 
the anisotropy increases the period from the precession time $2\pi/\hti$. The Toffoli gate for a given time $\tG$ can be accomplished by imposing the condition Eq.~\eqref{e.tg_cond} where
\begin{align}
 T_{01} =& \frac{2\pi}{\hti_{01}}~K\bigg(\frac a{\hti_{01}}\bigg)
         =  \pi~K\bigg(\frac a2\bigg)
\\
 T_{11} =& \frac{2\pi}{\hti_{11}}~K\bigg(\frac a{\hti_{11}}\bigg)
         = \frac{\pi}{2\cos\varphi}~K\bigg(\frac a{4\cos\varphi}\bigg)~,
\end{align}
so, choosing $m=n=1$ as above, one finds an implicit condition that determines $\varphi$,
\begin{equation}
 \frac3{4\cos\varphi}~K\bigg(\frac a{4\cos\varphi}\bigg)= K\bigg(\frac a2\bigg)~.
\end{equation}

\section{Concluding Remarks}
In this paper we have analysed several setups that realize a magnetic implementation of a Toffoli gate, under some suitable assumptions about the building elements of the device. Specifically, the control spins must be kept fixed during the target spin evolution. This requires that at least one of the following conditions to be satisfied: i) an external magnetic field can be applied to the target spin only [this is implicitly assumed by writing the Hamiltonian as in Eqs.(\ref{e.ham}) and (\ref{H.noncollinear})]; ii) the control spins are held along their initial orientation by stronger interactions with other circuit elements (e.g., as in Ref.\,\onlinecite{KhajetooriansWCW2011}); iii) a much higher single-ion anisotropy acts on the control spins. In the latter case we point out that, unless the anisotropy could be externally controlled on a time scale similar to the gate operation times, it may be difficult to make the scheme scalable, i.e., allowing for a given spin to act as a control or a target at different steps of the device 
operation.
 
Moreover, while the Hamiltonian model (\ref{e.ham}) allows for stationary logical states (with spins aligned along the quantization axis) for $h_\perp=0$, this does not hold for the non-collinear scheme, where timing thus becomes even more critical and it not only requires to have control over the external magnetic field, but also over the control-target interaction. Namely, one has to be able to switch on/off both the interaction and the field in very short times. Once this requirement is satisfied, outside of the time intervals of gate operation the target is subjected only to the effect of the anisotropy, which can therefore be even small to assure the stability of the logical states.

On the other hand, in the collinear model, a single ion anisotropy on the target spin is not only required to properly set the quantization axis, but it has also to be strong enough, i.e. $a>2$, to stabilize the spin configurations representing the logical states. As the proper spin-flip operation can be achieved only if $h_\perp>a$, measuring spin values in $\hbar$ units and writing the control-target coupling as $J\,(S_1^z{+}S_2^z)S^z$, the anisotropy term as $-AS_z^2$, and the Zeeman term as $-g_s\muB\vec{S}\cdot\vec{H}$, we may go back to physical units,
\[
 H_\|=\frac{2JS}{g_s\muB}
~,~~~~~~
 A \gtrsim 2J 
~,~~~~~~
 H_\perp > \frac{AS}{g_s\muB} \gtrsim \frac{2JS}{g_s\muB}=H_\|~~,
\]
to get a quantitative estimate of the relevant parameters. For the system presented in Ref.~\onlinecite{KhajetooriansWCW2011} we can gauge $J\simeq0.3$~meV~$\simeq 3.5$~K, an order of magnitude similar to that observed also in molecular magnets\cite{GatteschiSessoliVillain06,TimcoMW2013,TimcoEtAl}, typically $J\simeq1-20$~K. Therefore we may assume $J\sim 1$~K as a reasonable value for the control-target coupling; taking $S\sim 3-5$, we get $H_\|\sim 10$~T, $A\sim5$~K and $H_\perp\sim 50-100$~T. Because these are rather high values, the non-collinear scheme is a better candidate in view of an experimental realization of the device. From the above calculations, the order of magnitude for the gate operation time is in the range of 10-100 ps.

The necessary requirements on the field strength and on the timing can be made less strict due to the effect of a weak dissipative environment. We have shown in Sec.\ref{s.damping} that, although the small price that has to be paid in terms of gate-operation time and thermal energy loss, the environment is beneficial for implementing the device. Indeed we have shown that if the non-dissipative dynamics, while being not perfect, brings the target not too far from the wanted logical state configuration, the effect of dissipation is to eventually stabilize the target spin.

\begin{acknowledgements}
L.B.~A.C.~and~P.V~acknowledge financial support from PNRR MUR project PE0000023-NQSTI. 
P.V~declares to have worked inthe framework of the Convenzione Operativa between the Institute for Complex Systems of CNR and the Department ofPhysics and Astronomy of the University of Florence.
S.B. acknowledges EPSRC grants EP/R029075/1 and EP/X009467/1.
\end{acknowledgements}

\appendix 

\section{Analytical treatment of the dynamics with single-ion anisotropy} \label{app} 
We consider the setting presented in Sec.~\ref{s.aniso}. 
In order to make some analytical progresses we rewrite Eqs.~\eqref{e.mag_eom} as
\begin{eqnarray}
 \dot x &=& y (\hti+2az) ~,
\label{e.first}  
\\
  \dot y &=& h_\perp s^z - x (\hti+2az) ~,
\label{e.second}
\\
  \dot z &=& - h_\perp y ~,
 \label{e.third}
\end{eqnarray}
where
\begin{equation}
 \hti\equiv{h}_\|+s_1^z+s_2^z
\end{equation}  
is the overall effective field along $z$ including the interaction with the control spins. Feeding Eq.~\eqref{e.first} with~\eqref{e.third} we find
\begin{equation}
 h_\perp\dot x = - \dot z\, (\hti+2az)~,
\end{equation}  
which can be directly integrated leading to
\begin{equation}
 h_\perp x = r_\pm-\hti z -az^2~,
\label{e.sx_sol}
\end{equation}
where the initial conditions, $x(0)=0$ and $z(0)=\pm1$, have been used and  $r_\pm\equiv{a\pm\hti}$. Finally, using the  expression in Eq.~\eqref{e.second} we find an equation for $z(t)$ as
\begin{equation}
  \ddot z -\hti r_\pm +e_\pm z +3a\hti z^2 +2a^2z^3  = 0~,
\label{e.diffeq_zcomp}
\end{equation}
where $e_\pm\equiv{h_\perp^2+\hti^2-2ar_\pm}$.
This equation contains two nonlinear terms arising from the presence of anisotropy. A direct integration yields to 
\begin{equation}
\begin{aligned}
  {\dot z}^2 &=h_\perp^2-r_\pm^2 +2\hti r_\pm z -e_\pm z^2 -2a\hti z^3 -a^2z^4
\\
        &= (1{-}z^2)\big[e_\pm+a^2 (1{+}z^2)\big] -2\hti(1{\mp}z)\big[\hti-az(1{\pm}z)\big] ~.
\end{aligned}
\label{e.first_int}
\end{equation}
The r.h.s. of the latter equality can be interpreted as minus a potential energy, which is bounded since the support of $z\in[-1,1]$ is compact. In other words, the motion of $z(t)$ is periodic, similar to a particle in a potential well.  Eq.~\eqref{e.first_int} allows one to express the solution of the EoM (\ref{e.diffeq_zcomp}) in an implicit form (valid until a turning point),
\begin{equation}
  t=\mp\int\limits_{\pm1}^{z(t)} \frac{dz}
     {\sqrt{h_\perp^2-r_\pm^2+2\hti r_\pm z -e_\pm z^2-2a\hti z^3- a^2 z^4}} ~.
\label{e.impl_sol}
\end{equation}

%



\end{document}